# High-throughput lensless whole slide imaging via continuous height-varying modulation of tilted sensor

SHAOWEI JIANG,[1,4] CHENGFEI GUO,[1,4] PATRICK HU[2], DEREK HU[3], PENGMING SONG,[1] TIANBO WANG,[1] ZICHAO BIAN,[1] ZIBANG ZHANG,[1] AND GUOAN ZHENG[1,*]

[1]Department of Biomedical Engineering, University of Connecticut, Storrs, CT, 06269, USA
[2]Department of Computer Science, University of California Irvine, Irvine, CA, 92697, USA
[3]Amador Valley High School, Pleasanton, CA, 94566, USA
[4]These authors contributed equally to this work
*Corresponding author: guoan.zheng@uconn.edu

**We report a new lensless microscopy configuration by integrating the concepts of transverse translational ptychography and defocus multi-height phase retrieval. In this approach, we place a tilted image sensor under the specimen for linearly-increasing phase modulation along one lateral direction. Similar to the operation of ptychography, we laterally translate the specimen and acquire the diffraction images for reconstruction. Since the axial distance between the specimen and the sensor varies at different lateral positions, laterally translating the specimen effectively introduces defocus multi-height measurements while eliminating axial scanning. Lateral translation further introduces sub-pixel shift for pixel super-resolution imaging and naturally expands the field of view for rapid whole slide imaging. We show that the equivalent height variation can be precisely estimated from the lateral shift of the specimen, thereby addressing the challenge of precise axial positioning in conventional multi-height phase retrieval. Using a sensor with 1.67-µm pixel size, our low-cost and field-portable prototype can resolve 690-nm linewidth on the resolution target. We show that a whole slide image of a blood smear with a 120-mm² field of view can be acquired in 18 seconds. We also demonstrate accurate automatic white blood cell counting from the recovered image. The reported approach may provide a turnkey solution for addressing point-of-care- and telemedicine-related challenges.**

Phase information characterizes the optical delay accrued during propagation. Light detectors can only measure intensity variation of the light waves. Phase information is lost during the detection process. Measuring the phase of light waves often involves additional experimental complexity, typically by requiring light interference with a known field, in the process of holography [1]. Phase retrieval provides an alternative to measuring phase without requiring sophisticated interferometric setups [2]. It can be achieved by introducing different types of diversity measurements, including transverse translational diversity [3-6], defocus multi-height diversity [7-11], multi-wavelength diversity [12-14], nonlinear diversity [15], among others. The first two types of diversity measurements are of particular interest in this letter.

The strategy of using transverse translational diversity for phase retrieval is termed ptychography. In 1969, the concept was first introduced in electron microscopy to address the phase problem of crystallography [16]. The modern implementation of ptychography is shown in Fig. 1(a), where the specimen is laterally translated through a spatially confined probe beam and the diffraction patterns are recorded in the far-field. The diffraction measurements serve as the Fourier magnitude constraints in the reciprocal space. The confined probe beam limits the physical extent of the object for each measurement and serves as the support constraint in the real space [3]. It is also possible to swap the real and the reciprocal space in ptychography using a lens. Fourier ptychography is one example that implements ptychographic phase retrieval using a microscope setup [17, 18].

Defocus multi-height diversity, on the other hand, introduces different object-to-detector distances for diffraction data acquisition. In 1968, the concept was first introduced for electron microscopy, termed 'focus series reconstructions' [7]. In the optical region, it has been demonstrated in wavefront reconstruction [8, 9] and shows great potentials for lensless on-chip microscopy [10, 11]. A typical implementation of defocus multi-height phase retrieval is shown in Fig. 1(b), where the specimen is axially translated to different defocus distances for image acquisition. In the reconstruction process, the complex object solution is iteratively propagated to different defocus planes and the captured images are enforced as magnitude constraints.

In this letter, we report a new lensless microscopy configuration by integrating the concepts of transverse translational ptychography and defocus multi-height phase retrieval. As shown in Fig. 1(c), we place a tilted image sensor under the specimen for linearly-increasing phase modulation along one lateral direction. Similar to the operation of ptychography, we laterally translate the specimen and acquire the diffraction images for reconstruction. Since the axial distance between the specimen and the sensor varies at different lateral positions, laterally translating the specimen effectively introduces defocus multi-height measurements while eliminating axial scanning. Transverse translation further introduces sub-pixel shift for pixel super-resolution imaging and naturally expands the field of view for rapid whole slide imaging.

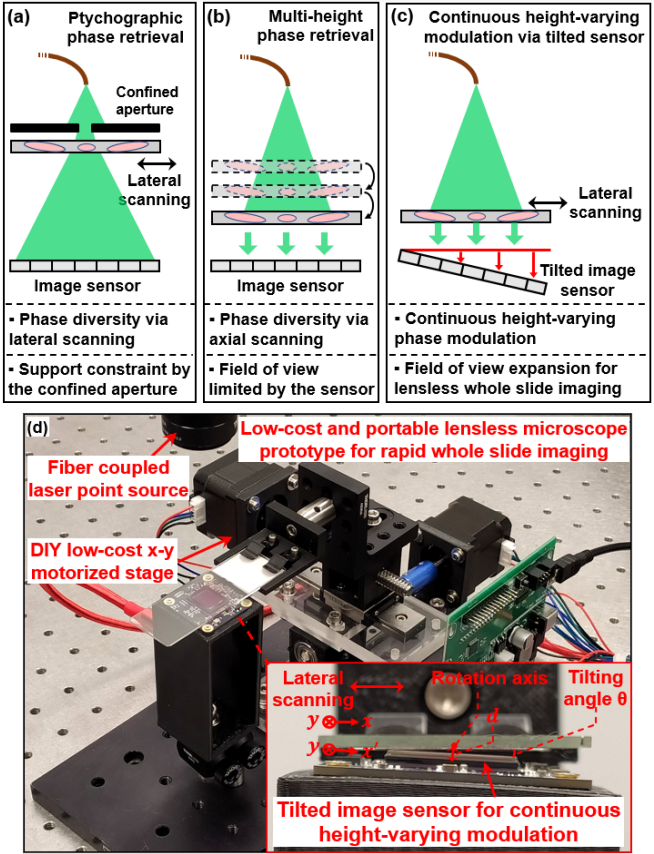

Fig. 1. Comparison between ptychography, multi-height phase retrieval and tilted sensor modulation. (a) Operation of ptychography, where the specimen is laterally translated through a confined aperture. (b) Multi-height phase retrieval, where the specimen is axially scanned to different defocus planes. (c) The reported lensless configuration where a tilted sensor is placed under the specimen for linearly increasing phase modulation along one lateral direction. Laterally translating the specimen effectively introduces defocus multi-height measurements while eliminating axial scanning. (d) The low-cost and portable prototype setup, with a 1.67-μm pixel size (Aptina MT9J003).

We summarize the advantages of the reported lensless configuration as follows. First, we do not rely on compact support constraint as that in ptychography. Instead, we employ full-field illumination as that in multi-height phase retrieval, enabling a large imaging field of view determined by the sensor area. Second, axial scanning of specimen or detector is a critical step in conventional multi-height measurement. The field of view is often limited by the area of the image sensor. The reported configuration can eliminate axial scanning and naturally expands the field of view. We show that a whole slide image of a blood smear with a 120-mm$^2$ field of view can be acquired in 18 seconds. Third, we show that the equivalent height variation can be precisely estimated from the lateral shift of the specimen, thereby addressing the challenge of precise axial positioning in conventional multi-height phase retrieval. Fourth, we employ a sub-pixel recovering model used in ptychography for bypassing the resolution limit imposed by the pixel size [5, 6, 19]. Using a sensor with 1.67-μm pixel size, our low-cost prototype can resolve 690-nm linewidth on the resolution target.

Our prototype platform is shown in Fig. 1(d), where specimen is mounted on a x-y motorized stage modified from a low-cost computer numerical control (CNC) router [20]. We couple a 532-nm laser diode to a single-mode fiber for sample illumination. The tilt angle between the specimen and the detector is 5 degrees. For a larger tilt angle, the distance between the sample and the sensor becomes larger and leads to a lower resolution. For a smaller tilt angle, the lateral translation in our setting leads to a smaller axial shift of the sample and it needs more measurements to perform the diversity-based phase retrieval. A 5-degree tilt angle is a good comprise between these two considerations. The forward imaging model of our system can be expressed as

$$I_j(x',y) = \left| tilt_\theta \left( O(x - x_j, y - y_j) \right) * PSF_{free}(d) \right|^2_{\downarrow M}, \quad (1)$$

where $I_j(x',y)$ is the $j^{th}$ intensity measurement at the tilted image sensor in Fig. 1(d), $O(x,y)$ is the complex exit wavefront of the object, $(x_j, y_j)$ is the $j^{th}$ lateral shift of the object, $PSF_{free}(d)$ is the point spread function (PSF) for free space propagation over a distance of $d$, '*' denotes the convolution operation. We use '$\downarrow M$' to represent the down-sampling process ($M = 3$). In Eq. (1), we use $tilt_\theta$ to model the tilting process of the object wavefront at the $(x, y)$ plane to the $(x', y)$ plane, with a tilt angle of $\theta$. The tilting axis is at the central column of the object wavefront.

---

Algorithm outline

Input: Raw images $I_j$ ($j = 1, 2, \cdots, J$), the tilting angle $\theta$, and the initial distance $d$ between the object and the image sensor
Output: High-resolution object $O(x, y)$
---

1    Calculate the translational shift $(x_j, y_j)$ of the object via cross-correlation
2    Initialize the tilted object wavefront $W(x', y)$ parallel to the image sensor
3    for $n = 1: N$ (different iterations)
4      for $j = 1: J$ (different captured images)
5        $d_j = tan(\theta) \cdot \sqrt{x_j^2 + y_j^2}$    % Calculate the equivalent height variation
6        $W_j(x', y) = W(x' - x_j, y - y_j)$    % Shift the wavefront to $(x_j, y_j)$ position
7        $W_{j,crop}(x', y) = W_j(x_l': x_h', y_l: y_h)$    % Crop the central region
8        $\psi_j(x', y) = PSF_{free}(d + d_j) * W_{j,crop}(x', y)$    % Light on the sensor plane
9        Update $\psi_j(x', y)$ using Eq. (2)    % Up-sampled magnitude projection
10       $W'_{j,crop}(x', y) = PSF_{free}(-(d + d_j)) * \psi'_j(x', y)$    % Propagate it back
11       $W_j(x_l': x_h', y_l: y_h) = W'_{j,crop}(x', y)$    % Update the central region
12       $W(x', y) = W_j(x' + x_j, y + y_j)$    % Shift back the updated wavefront
13     end
14   end
15   $O(x, y) = tilt_{-\theta}(W(x', y))$    % Tilting back the updated object wavefront

Fig. 2. Reconstruction process of the reported platform.

Figure 2 shows the reconstruction process of the reported platform. This process aims to recover the exit wavefront of the object $W$ at the tilted plane $(x', y)$ first and then recover the object $O$ at the horizontal plane $(x, y)$. The tilt angle can be estimated by propagating two different regions of the captured images to two different in-focus positions. The tilt angle can then be calculated based on the height difference and the lateral distance of the two regions. Similarly, the distance between the sensor and the sample can be estimated post-measurement via an autofocusing metric [21]. The translation positions can be estimated via cross-correlation analysis [22]. In line 5 of Fig. 2, we estimate the equivalent height variation from the lateral shift of the specimen. In line 9, we employ the following up-sampling process to perform pixel super-resolution imaging [5, 6, 19]:

$$\psi'_j(x',y) = \psi_j(x',y) \left( \frac{\sqrt{I_j(\lceil x'/M \rceil, \lceil y/M \rceil)}}{\sqrt{U_j(\lceil x'/M \rceil, \lceil y/M \rceil)}} \right), \quad (2)$$

where '$\lceil \ \rceil$' represents the ceiling function and $U_j(x',y) = |\psi_j(x',y)|^2 * ones(M,M)$. Equation (2) essentially enforces that the intensity summation of every $M$ by $M$ sub-pixels in the complex wavefront is equal to the corresponding raw pixel in the captured image.

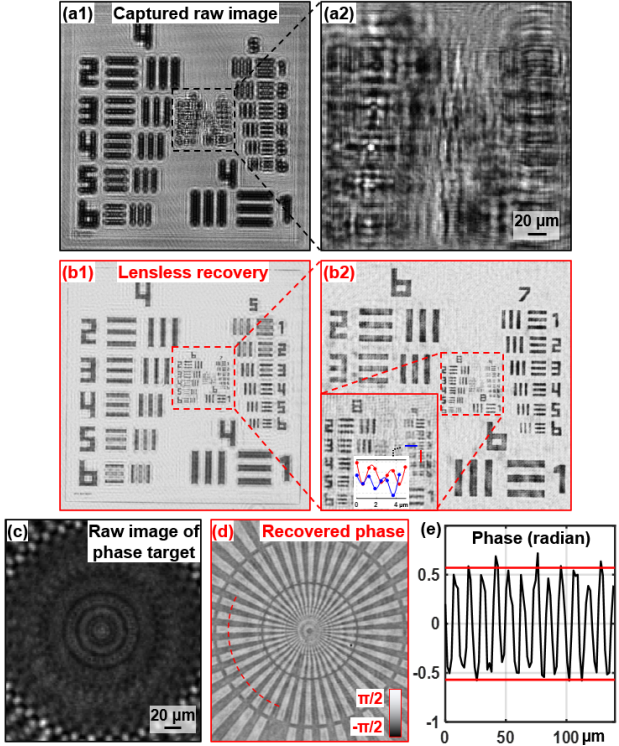

Fig. 3. Validation using a resolution target and a phase target. (a) The captured raw image of the resolution target. (b) The reconstruction for the resolution target based on 10 measurements. (c) The raw image of the phase target. (d) The recovered phase. (e) The line trace along the dashed line in (d).

The tilting process in line 15 of Fig. 2 can be implemented by applying a spectrum interpolation in the Fourier domain. In our implementation, we tilt the sensor along the y-axis (i.e., y-axis rotation). The corresponding inverse transformation matrix $T^{-1}$ from the plane of $(x',y)$ to the plane of $(x,y)$ can be written as follows:

$$T^{-1} = \begin{bmatrix} \cos(\theta) & 0 & -\sin(\theta) \\ 0 & 1 & 0 \\ \sin(\theta) & 0 & \cos(\theta) \end{bmatrix}, \quad (3)$$

We can then define the following spectrum of the object $\hat{O}$ and the Jacobian $J$ based on the weights of $T^{-1}$:

$$\hat{O}(k_x, k_y) = \widehat{W}(k'_x \cos(\theta) - k'_z \sin(\theta), k_y), \quad (4)$$

$$J(k'_x, k_y) = \cos(\theta) + \frac{k'_x}{k'_z} \sin(\theta), \quad (5)$$

where $\widehat{W}(k'_x, k_y)$ is the spectrum of the recovered tilted wavefront in line 12 of Fig. 2 and $k'_z = (k_0^2 - k_x'^2 - k_y^2)^{1/2}$ ($k_0$ is the wave number in vacuum). The refocused object at the horizontal plane $(x,y)$ can be obtained via

$$O(x,y) = \mathcal{F}^{-1}\{\hat{O}(k_x,k_y) \cdot |J(k'_x,k_y)|\} \quad (6)$$

We first validate our platform using a resolution target and a quantitative phase target in Fig. 3. In this experiment, we translate the sample to 10 different lateral positions with a lateral step size of 30-50 μm. The reconstruction time is ~80 seconds for processing 10 raw measurements. Figure 3(a) shows the captured raw image of the resolution target and Fig. 3(b) shows the reconstruction, where we can resolve the 0.69-μm linewidth from group 9, element 4. We note that the achieved resolution has a small difference in the two orthogonal directions. It is possible to achieve isotropic resolution by scanning / tilting the sample in two directions. To further improve the resolution, we need to model the pixel PSF and properly deconvolve it in the reconstruction. Using an image sensor with a smaller pixel size is another straightforward solution to improve the resolution. Figure 3(c) shows the captured raw image of the phase target. Figure 3(d) shows the recovered phase. As shown in Fig. 3(e), the recovered phase is in a good agreement with the ground-truth of the phase target (two red lines).

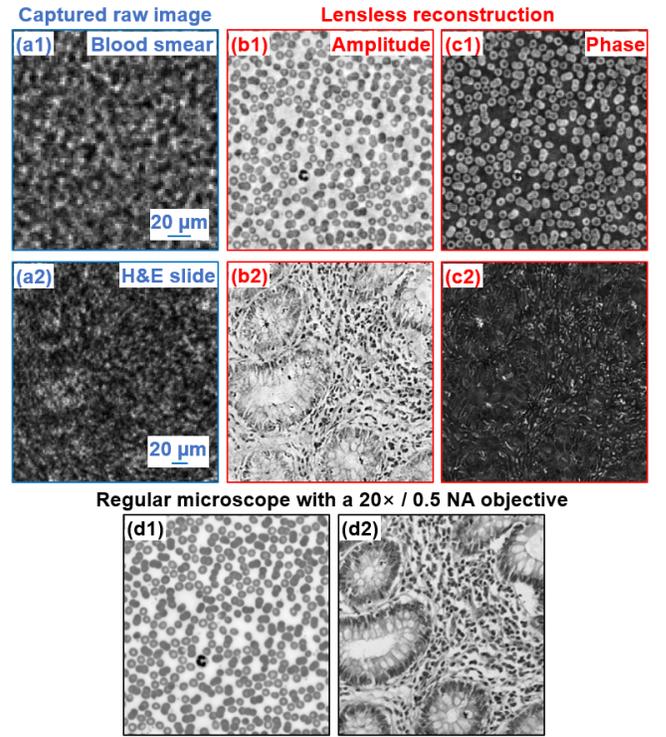

Fig. 4. (a) Captured raw images of three different biological specimens. (b) Recovered amplitude and (c) phase of the specimens using 10 raw measurements. (d) The images captured using a 20× / 0.5 NA lens.

In the second experiment, we test our platform using a blood smear and a hematoxylin and eosin (H&E) stained esophagus cancer slide. Figure 4(a) shows the captured raw images of the three samples. Figures 4(b) and 4(c) show the recovered amplitude and phase using 10 raw measurements. Figure 4(d) shows the images captured using a 20× lens.

In the third experiment, we demonstrate the application of the reported platform for high-throughput whole slide imaging of a blood smear and perform automatic white blood cell counting. Figure 5(a) shows the whole slide image of the blood smear with a field of view of ~120 mm². This image contains 4 field of views of the detector size (40 raw images) and the acquisition time is ~18 seconds. Based on the lensless recovery, we perform virtual staining [23] to generate the color whole slide image in Fig. 5. Figure 5(b) shows the zoom-in views of Fig. 5(a) and Fig. 5(c) shows the images captured using a 20× objective lens. We also perform automatic white blood cell counting based on the whole slide image. 1014 white blood cells have been automatically located from whole slide image. The inset of Fig. 5(a) shows the locations of these white blood cells. The number of white blood cells exactly matches the manual counting result.

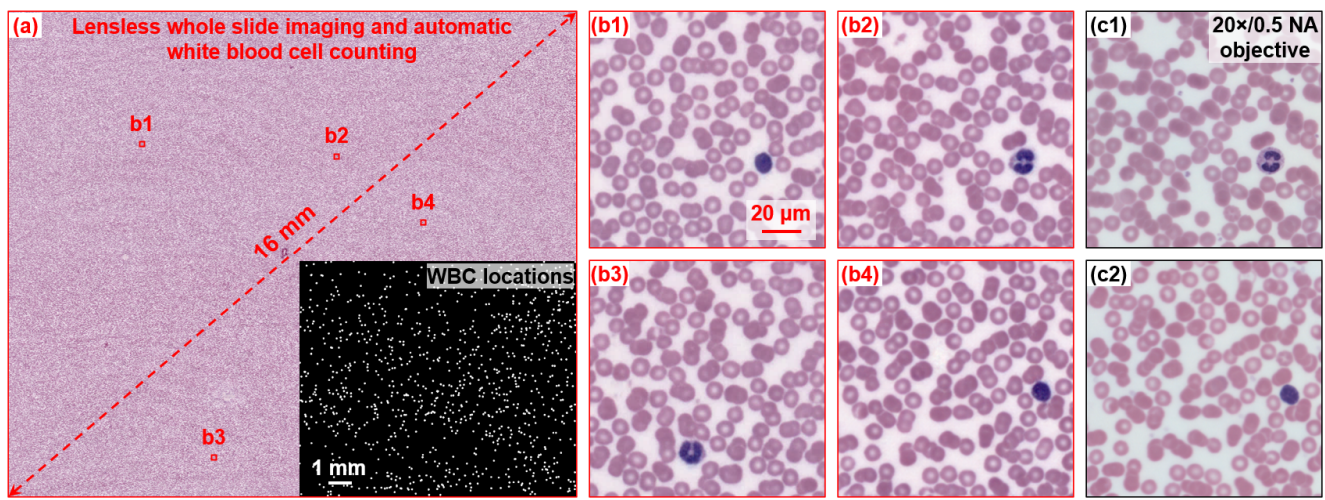

Fig. 5. Lensless whole slide imaging of a blood smear. (a) The virtually stained image based on the lensless recovery using the reported imaging platform. Inset shows the locations of the tracked 1014 white blood cells. The acquisition time is 18 seconds with a field of view of ~120 mm$^2$. (b1)-(b4) Zoom-in views of (a). (c) The images captured using a 20× / 0.5 NA objective lens.

In summary, we report a new lensless configuration by integrating the concepts of transverse translational ptychography and defocus multi-height phase retrieval. By placing a tilted image sensor under the specimen, we introduce linearly increasing phase modulation along one lateral direction. Laterally translating the specimen in our platform, therefore, effectively introduce defocus multi-height measurements while eliminating axial scanning. We show that the equivalent height variation can be precisely estimated from the lateral shift of the specimen, thereby addressing the challenge of precise axial positioning in conventional multi-height phase retrieval. We also show that a whole slide image of a blood smear with a 120-mm$^2$ field of view can be acquired in 18 seconds. One future direction is to integrate multi-angle illumination in Fourier ptychography for synthetic aperture imaging and diffraction tomographic imaging.

**Funding.** National Science Foundation 1700941 and 2012140.

**Disclosures.** The authors declare no conflicts of interest.